\documentclass[twocolumn]{aastex62}
\usepackage{amsmath,amssymb,amstext}
\usepackage{gensymb}
\usepackage{natbib}
\shorttitle{Implications of a Surface}
\shortauthors{May \& Rauscher}

\begin{document}
\title{From Super-Earths to Mini-Neptunes: Implications of a Surface on Atmospheric Circulation}

\correspondingauthor{E. M. May}
\email{erin.may@jhuapl.edu}
\author{E. M. May}
\affil{Johns Hopkins Applied Physics Laboratory \\
Laurel, MD}
\affil{University of Michigan \\
Ann Arbor, MI}
\author{E. Rauscher}
\affil{University of Michigan \\
Ann Arbor, MI}
\begin{abstract}
It is well known that planets with radii between that of Earth and Neptune have been the most commonly detected to-date. To classify these planets as either terrestrial or gaseous, typically we turn to mass-radius relations and composition curves to determine the likelihood of such a planet being rocky or gaseous. While these methods have set a likely transition radius of approximately 1.5 R$_{\oplus}$, we cannot expect that any change between terrestrial and gaseous compositions will be a sharp cut-off, and composition curve predictions result in ambiguous designations for planets right near this transition radius. In this work we present 3D general circulation models of transition planets, wherein we study the effects of a surface on observable quantities such as the latitudinal variations and eclipse depths. We present our updated GCM, validated on the circulation of Earth, before discussing our modeling choices for this transition planet. Finally, we discuss the results of this study and explore the prospects of detecting the presence of a surface through observations of secondary eclipses in the future.
\end{abstract}
%
%
\section{Introduction}
From exoplanet detections to-date, planets with radii between that of Earth and Neptune (3.88R$_{\oplus}$, the smallest gaseous planet in our Solar System) have been discovered to be the most common type of planet outside of our Solar System \citep{Batalha2014}. Naturally, there must be some point between these two radii at which we see a transition from primarily terrestrial planets to primarily gaseous planets. Without a direct comparison in our Solar System, we must turn to other methods to understand the unique properties of this class of planets. 

This transition from terrestrial to gaseous has been studied by \cite{Rogers2015} which sets the transition from terrestrial to gaseous at 1.6R$_{\oplus}$ and \cite{LopezFortney2014} which sets the transition at 1.75 R$_{\oplus}$. In addition, \cite{Fulton2017} find a gap in the known planet radius distribution at 1.5 R$_{\oplus}$. While these two populations could be formed in a variety of ways, possibly due to photoevaporation rather than a result of formation, they are concrete evidence for the existence of separate, yet overlapping, radii regimes for Super-Earths and Mini-Neptunes. In this work, the determination of if a planet is currently terrestrial or gaseous is only concerned with the current presence of a surface and not necessarily how and/or why it got to where it is now. As calculated in \cite{LopezFortney2014}, there is unlikely to be a way to make a planet larger than 2.0 R$_{\oplus}$ without a significant gaseous envelope. This is an important distinction when classifying planets as terrestrial or gaseous near the radius gap.

Because mass and radius are commonly the two most important factors differentiating `types' of exoplanets as we currently define them, an active area of research is determining a robust relation between the two. Such a relation is useful in determining the composition of an exoplanet, due to density generally being a strong indicator of composition. \cite{Wolfgang2016,Ning2018} and \cite{Kanodia2019} are examples of mass-radius relations tuned for various types of exoplanets, with a focus on the predictive power of such techniques to determine a likely mass or radius if one is not known.

However, even when a mass and radius are known for a transition regime planet, uncertainties on these values, the wide number of possible compositions that match its derived density, and observed and expected scatter in mass-radius relations leave some ambiguity in its likelihood to be gaseous or terrestrial. Such compositions curves are presented in \cite{Seager2007} and \cite{Zeng2016} for various commonly considered compositions including Iron, MgSiO$_3$ (rock), H$_2$O, in addition to combined multi-layer compositions. One can add more composition curves by including further combinations of compositions \citep[][do this in their computer readable tables]{Zeng2016} and by including H$_2$/He envelopes of varying mass-fractions. 

Naturally there exists more than one composition curve that matches any given measured or inferred planetary density including options that would classify a planet as both terrestrial and gaseous over this intermediate-size regime. While the various composition curves that match a given exoplanet will lead to composition differences that may be detected through atmospheric observations, the likely presence of aerosols in the planet's atmosphere will inhibit our ability to make a direct determination of composition for all planets. For example, GJ 1214b, while unlikely to be terrestrial due to it's radius of 2.678 R$_{\oplus}$ \citep{Charbonneau2009}, appears cloudy out through near-infrared wavelengths as observed with the Hubble Space Telescope \citep{Kreidberg2014}, precluding our ability to directly measure the composition of the atmosphere. Because this is among the smallest planets for which we've been able to directly probe the atmosphere, it leads to reasonably uncertainity as to our ability to use direct measurements of composition to make statements about a planet's liklihood to be terrestrial or gaseous.

For this reason, we seek to determine a density-independent classification scheme for transition regime planets by considering the circulation effects of having a solid surface. Terrestrial and gaseous circulation patterns have been well studied independently; see, among others, \cite{Showman2013,KS15,Koll2016,Way2018,Komacek2019,Komacek2019b,Pier2019} for terrestrial examples and \cite{Perez2013,Rauscher2014,Showman2015,Kataria2016,Komacek2016,Rauscher2017} for gaseous examples. Each of these looks at variations in specific key parameters that affect circulation for either terrestrial or gaseous planets independently. Here we present a single modeling scheme applied uniformly for Super-Earths and Mini-Neptunes alike. Our model is bench-marked against Earth, a planet for which we know the circulation and emission well.

Recent work by \cite{Kreidberg2019} presented a phase curve of the small 1.3 R$_{\oplus}$ planet LHS 3844b as a method to detect the presence (or absence) of an atmosphere. The team found that the object was likely a rocky body without any significant atmosphere, the first confirmation of a rocky or gaseous composition outside of mass-radius constraints alone. While this method is extremely promising for classification of these transition planets, it is also observationally expensive owing to the need to continuously observe the target over the course of an entire orbit to obtain full phase coverage. This limits the technique to only planets on sufficiently short orbits such that they are observationally feasible, and regardless of the orbital period, the large duration of observation required in combination with limited telescope time limits the number of planets that can be studied in this manner. In this work we focus on alternative classification methods using only secondary eclipses and what we can learn from eclipse mapping. 

The recent set of papers by \cite{Mansfield2019, Koll2019, Malik2019} also look at the possibility of using secondary eclipses to determine the presence of an atmosphere for tidally locked terrestrial planets and suggest that a detected albedo of at least 0.5-0.7 is high enough to differentiate between surface reflection and high altitude clouds. Here we explore the possibility of detecting atmospheres of non-tidally locked terrestrial planets without clouds, where surface and top of the atmosphere reflection is not easily disentangled. In this case, the key feature of a surface will, instead of albedo, be related to the heat redistribution in the atmosphere as the surface helps to more effectively move heat away from the equator.

In the coming years, the launch of the James Webb Space Telescope (JWST) will enable eclipse mapping of Hot Jupiters down through even a temperate Earth-sized planet in over a dozen eclipse observations \citep{Belu2011,Beichman2014,Schlawin2018}. Eclipse mapping enables the determination of longitudinally and latitudinally resolved maps of the planet's emission and/or reflection \citep{Williams2006,Rauscher2007}. The shape of ingress and egress during secondary eclipse determines a map of the planet's emission by measuring how much flux is blocked by the host star from successive arcs of the planet as it moves out of and back into our line of sight. Recently, \cite{Rauscher2018} demonstrated a new method to more accurately retrieve maps in the era of JWST. In this work, we focus on the zonally averaged latitudinal effects of a surface on the atmospheric circulation, and the observational consequences of surface in the context of the era of JWST. For non-tidally locked (quickly rotating) planets, this equator to pole heat transport, and observability of the resulting temperature gradient, is the main surface effect we quantify in this work.

In Section \ref{Surf:methods} we discuss our modeling framework including the general circulation model applied and updates to replicate the effects of a surface. In Section \ref{Surf:validation} we present our results for our Earth model as a validation of our modeling framework. Section \ref{Surf:TransitionPlanet} outlines our modeling choices for our transition planet. In Section \ref{Surf:TPResults} we present our results and the observational consequences of a surface. 

\section{Method} \label{Surf:methods}
%
\subsection{General Circulation Model}
We employ a three dimensional general circulation model (GCM), a robust tool in modeling planetary atmospheres. Our GCM is outlined in detail in \cite{RM12} (and henceforth referred to as RM12), with modifications as discussed in the following sections. The GCM is built upon the primitive equations of meteorology, a standard reduction of the Navier-Stokes equations under assumptions of an inviscid flow, vertical hydrostatic equilibrium, and small relative vertical flow and scales. \cite{Vallis2006} contains an in-depth derivation and discussion of these equations and assumptions. RM12 uses a double-grey approximation (visible band for stellar irradiation, infrared band for emitted heat), with the radiative transfer recently updated as described in \cite{Roman17}, following the scheme of \cite{Toon89}. 

RM12 originated as an Earth-Based code, using a dynamical core developed at the University of Reading \citep{Hoskins1975} and as such has been well tested for terrestrial planet applications \citep{Joshi1995,deForster2000,Menou2009}. Having undergone major changes for applications to gaseous exoplanets, in this work we re-introduce a surface into the GCM, including the relevant interactions between the surface and the atmosphere. To date, RM12 has been used to study, among other things, observational signatures of non-synchronous rotation \citep{Rauscher2014}, atmospheric circulation of circumbinary planets \citep{May2016}, observational signatures of obliquity \citep{Rauscher2017}, the radiative effects of clouds \citep{Roman17,Roman2018}, and to constrain high-resolution spectroscopic observations \citep{Zhang2017,Flowers2018}.
%
\subsection{Surface-Atmosphere Interactions}
We include two main surface-atmosphere interactions as outlined below. First, we consider the additional heating sources due to the surface. Second, we consider the effects of drag on the atmosphere. 
\subsubsection{Surface Heating}
The introduction of a surface requires us to consider the additional heating sources that naturally will result. Our bottom boundary flux condition is no longer the planet's internal heat source, but is rather determined by emission and/or reflection from the surface. Each surface element below an atmospheric column is treated independently, i.e. there is no heat transport within the surface to neighboring resolution elements. This choice allows us to apply the one-dimensional heat equation to determine the surface temperature. We do not consider moist effects in this work, and as such do not include latent heating terms. Therefore, the equation governing the energy exchange with the surface is given by
\begin{multline}
c_{s} \rho_{s} \Delta z_{s} \frac{\partial T_{s}}{\partial t}= \left(1-\alpha_{SW}\right)F_{SW\downarrow}+\left(1-\alpha_{LW}\right)F_{LW\downarrow}\\-j_{s}\sigma T_{s}^{4}-c_{a}\rho_{a} C \left|\overrightarrow{u_{a}}\right|\left(T_{a}-T_{s}\right)
\label{eqtn:surface_heating}
\end{multline}
where $c_{s}$ is the surface specific heat; $\rho_{s}$ is the surface density; $\Delta z_{s}$ is the thickness of the surface layer; $T_{s}$ is the surface temperature; $F_{SW\downarrow}$ and $F_{LW\downarrow}$ are the downwards short wave (optical) and long wave (infrared) fluxes, respectively; $\alpha_{SW}$ and $\alpha_{LW}$ are the short wave and long wave surface albedos, respectively; $j_s$ is the surface emissivity; $\sigma$ is the Stefan-Boltzman constant;  $c_{a}$ is the specific heat of the atmosphere; $\rho_{a}$ is the density of the atmospheric level directly above the surface; $C$ is the transfer coefficient for specific heat, set to 10$^{-3}$ following estimates in \cite{Frierson2006}; $\overrightarrow{u_{a}}$ is the wind vector in the atmospheric level directly above the surface; and $T_{a}$ is the temperature of the atmospheric level directly above the surface. This last term describes the sensible heat, which is an energy exchange as winds blow across the surface layer.  This surface layer scheme is similar to that used by \cite{KS15} and \cite{Frierson2006}, however we choose to use a uniform solid surface, rather than a water slab. The choice of a solid surface rather than a water slab is made to eliminate moist effects (ocean evaporation and cloud formation) from our model, which are beyond the scope of this work.

The surface emission plus reflection off the surface in the long wave is treated as the long wave (infrared) bottom boundary condition for the atmosphere; with the short wave reflection the bottom boundary condition in the short wave (visible), for an optically thin atmosphere. This results in an additional source of heating, working to warm the atmosphere from below. In our standard RM12 GCM, the bottom boundary condition takes the downwards long wave flux and adds it into the upwards long wave flux in the bottom layer of the atmosphere to maintain energy conservation. With the inclusion of this new physically motivated boundary condition, our lowest atmospheric levels are considered realistic representations of the dynamics that occur there.
\subsubsection{Atmospheric Drag}
We introduce drag into the atmosphere through Rayleigh Friction, following the benchmark work by \cite{HS94}. The friction is a decaying function of pressure, with the strongest effect in the atmospheric level directly above the surface. Rayleigh friction is given by
\begin{equation}
\label{Surf:surfaceheat_eq}
\frac{\partial \left|\overrightarrow{u_{a}}\right|}{\partial t}\left(\sigma\right)=-k_f \mathrm{max}\left[0,\frac{\sigma-\sigma_{b}}{1-\sigma_{b}}\right] \left|\overrightarrow{u_{a}}\right|
\end{equation}
where $\overrightarrow{u_{a}}$ is the wind speed in the given $\sigma$ level; $k_{f}$ is the coefficient of friction, in units of the inverse of time; $\sigma$ corresponds to the atmospheric level, with $\sigma_{b}$ being the boundary above which friction no longer exists (sometimes referred to as the boundary level). The $\sigma$ levels are a common realization of pressure in GCMs, where $\sigma\equiv P/P_0$ with $P_0$ the reference pressure, i.e. the average pressure at the bottom level of the model. Following \cite{HS94} and their Earth models, we set $k_{f}$=1 day$^{-1}$ and $\sigma_{b}$=0.7 to recreate an Earth-like friction profile. 
\par Although it is a small fraction of the total energy budget, in order to maintain energy conservation, we choose to return the energy removed in a given cell through drag directly back to the atmosphere as heating in the same cell. The temperature increase due to returned energy is given by
\begin{equation}
\label{Surf:rayfric_eq}
\Delta T_{a}=\frac{\Delta \left|\overrightarrow{u_{a}}\right|^2}{2c_{a}}
\end{equation}
where all terms are the same as above. The general result of drag is to slow the winds near the surface, and add a small amount of additional heating to the atmosphere. Because the friction in our model is set up to match the prescription of \cite{HS94}, our choice of a solid surface vs. a slab ocean does not affect the strength of the friction we apply on the deeper levels of the atmosphere.

\subsection{Choice of Model Resolution}

Baroclinic instabilities are a circulation pattern consisting of eddies that arises due to the rotation of the atmosphere and the missalignment of the vertical stratification with the density gradient. Typically, there are between 5 and 8 large scale eddies that form around the Earth at mid latitudes which result in efficient transport of heat pole wards in these regions compared to the transport provided by the mean flow. For a more in depth discussion of the effects of these eddies on circulation, see \cite{WPBook} and \cite{HoltonBook}.

The scale of the baroclinic eddies is typically  proportional to the Rossby radius of Deformation, given by
\begin{equation} \label{Surf:eqtn:rrd}
    L=\frac{NH}{\pi f}
\end{equation}
where $H$ is the scale height; $f$ is the coriolis parameter given by $f=2\Omega\sin{\gamma}$ with $\Omega$ the rotation rate and $\gamma$ the latitude; and $N$ is the Brunt-V\"{a}is\"{a}l\"{a} frequency given by
\begin{equation} \label{Surf:eqtn:bvf}
    N=\sqrt{-\frac{g}{\theta}\frac{d\theta}{d z}}
\end{equation}
where $g$ is the gravity; $z$ is the vertical unit of height; and $\theta$ is potential temperature given by 
\begin{equation} \label{Surf:eqtn:ptp}
    \theta=T\left(\frac{P_0}{P}\right)^{R/c_p}
\end{equation}
with $T$ the temperature; $P$ the pressure and $P_0$ the reference pressure; $R$ the ideal gas constant; and $c_p$ the specific heat capacity of the atmosphere. We test various horizontal resolutions around the expected necessary resolution to insure we have sufficiently resolved these important dynamical features in the circulation. Since we are presenting models for planets with different dynamical scales, we discuss the specific horizontal resolution choices below.

\section{Model Validation - Earth} \label{Surf:validation}
First, we run models for Earth to validate the updated RM12 GCM on terrestrial planets. Because we have direct observations of Earth's circulation, we are able to compare our models to data and previous modeling work.

In all runs which include a solid surface boundary, we treat the surface as an uniform composition slab composed of enstatite (MgSiO$_3$). Common Earth-like compositions in mass-radius relations for rocky exoplanets include 67.5\% enstatite by weight, with a pure-enstatite body representing the ``pure-rock'' composition \citep{Zeng2019} . Though the heat capacity of the surface ($c_s$  in Equation \ref{eqtn:surface_heating}) is itself a function of temperature \citep{Krupka1985}, we choose to hold this constant to minimize variations in parameters, 
and set $c_s$ to a value corresponding to a temperate of 500K which represents the average surface temperature across our models. Table \ref{Surf:surfparam} summarizes the input parameters for our solid surface. 

We use a diurnally averaged heating scheme for all models, in which the stellar irradiation pattern is applied in a daily average around the entire globe such that the equator receives more irradiation that the poles. This is an obvious choice for Earth and Neptune based on their observed insolation patterns and rotation rates.

\begin{deluxetable}{lr} 
\tablecaption{Surface Heat Equation Parameters \label{Surf:surfparam}}
\tablehead{
	\colhead{Parameter}		&
	\colhead{Value Adopted}			           	}
\startdata
            c$_s$  (heat capacity)     &   528.99 J kg$^{-1}$ mol$^{-1}$   \\
            $\rho_s$  (surface density)  &   3.5 g cm$^{-3}$                 \\
            $\Delta z$ (surface thickness) &   100 cm                             \\
            $j_{s}$ (surface reflectively)   &   1.0     	                        \\
            A (surface albedo)) &   0.3                             \\
\enddata
\tablecomments{See Equation \ref{Surf:surfaceheat_eq} and following text. Heat capacity (c$_s$) is calculated following \cite{Krupka1985} at a temperature of 500K. Emissivity ($j_s$) is an average infrared value derived from \cite{Mat2016} and for modeling stability. Our surface layer is chosen to be 1 meter thick as in \cite{KS15}.}
\end{deluxetable}
%
As in \cite{HS94}, we run multiple Earth-like models at resolutions of T21, T42, and T63 to study the resolution effects on the induction of baroclinic instabilities in the atmosphere due to the surface-atmosphere interactions. The necessary resolution can be estimated based on the scale of the Rossby radius of deformation, given in equations \ref{Surf:eqtn:rrd} - \ref{Surf:eqtn:ptp}. On Earth, $L$ at mid-latitudes is $\sim$1000 km. In order to sufficiently resolve these features, we need our resolution to be at least half this size. For Earth, a T21 resolution corresponds to a resolution at the equator of 625 km, T42 corresponds to 310 km, and T63 corresponds to 210 km. We therefore expect that the T21 run might not be high enough resolution to fully resolve and excite the baroclinic instability features. 

All Earth runs are done in linear pressure space. Typically, log pressure are used for gas giants with thick atmospheres in order to more highly resolve low-pressures where the main heating and cooling of the atmosphere driving the circulation takes place. For planets with thin atmospheres, and particularly in those cases with a surface, we are more interested in resolving regions near the surface where the key interactions occur, and so linear pressure levels makes more sense. All Earth models cover 0 - 1 bar linearly over 30 sigma-levels.
\subsection{Model Inputs}
The following discussion of model inputs for Earth is summarized in Table \ref{Surf:earthparam}. Most important to the resulting heating profile is choosing absorption coefficients to correctly represent the observed temperature-pressure profile on Earth. We model only the troposphere and assume the temperature would continue to fall with decreasing pressure without the turnover at the tropopause (on Earth, this is a result of the ozone present in the stratosphere).

\begin{deluxetable}{lr} 
\tablecaption{Earth GCM Input Parameters \label{Surf:earthparam}}
\tablehead{
	\colhead{Parameter}		&
	\colhead{Value Adopted}			           	}
\startdata
            g (surface gravity)              &       981 cm s$^{-2}$                        \\
            R (specific gas constant)   &       287 J kg$^{-1}$ K$^{-1}$        \\
            R$_{\oplus}$  (planet radius)  &       6.371$\times$10$^{8}$ cm                    \\
            $\Omega$ (planet rotation rate)      &       7.292$\times$10$^{-5}$ radians s$^{-1}$     \\
            P$_0$  (surface pressure)         &       1 bar                                       \\
            F$_{int}$ (internal heat)      &       0.087 W m$^{-2}$                            \\
            F$_{irr}$ (stellar irradiation)     &       1370 W m$^{-2}$                             \\
            A (TOA albedo)           &       0.3                       \\
            $\kappa_{v}$ (visible absorption coefficient)   &       2.45$\times$10$^{-4}$ cm$^{2}$ g$^{-1}$     \\
            $\kappa_{th}$ (thermal absorption coefficient)  &       3.50$\times$10$^{-3}$ cm$^{2}$ g$^{-1}$     
\enddata
\end{deluxetable}

We first consider the averaged optical depth of Earth's atmosphere in the optical band to set the photospheric level. For the shortwave (visible) band, we take this band to encompass all incoming stellar irradiation. We use that $\sim$ 23\% of incoming solar irradiation is absorbed by the atmosphere \citep{NEO} to approximate the optical depth of Earth's atmosphere in the shortwave band as 0.25 at 1 bar where $\tau$=$\kappa_{sw}P/g$ \citep[for double gray radiative formulation, see][]{Guillot2010}. This gives us an absorption coefficient of $\kappa_{sw}$=2.45$\times$10$^{-4}$cm$^2$ g$^{-1}$ and a photosphere pressure of p$_{sw}$=4.00 bar. 

To set the absorption in the infrared, we assume $\gamma$=0.07  \citep[commonly used for the super-Earth GJ 1214b, see][]{MR&F10,Miguel2014} where $\gamma$=$\kappa_{sw}/\kappa_{lw}$. This corresponds to an absorption coefficient of $\kappa_{lw}$=3.5$\times$10$^{-3}$ cm$^2$ g$^{-1}$, with a photosphere pressure of p$_{lw}$=0.28 bar. We find that these values reproduce the shape of Earth's temperature-pressure profile well, but result in higher temperatures than those calculated with the \cite{Guillot2010} profile. We explain this with the lack of surface interactions in the standard \citeauthor{Guillot2010} profiles which results in excess heating in the lower atmosphere. 

With the above absorption coefficients, our Earth model has a surface which is above the short wave (visible) photosphere but below the long wave (infrared) photosphere. In reality, the long wave absorption on Earth is slightly more complex than the short wave since Earth's absorption spectrum is highly variable at infrared wavelengths with several windows in which the atmosphere is transparent and several at which it is opaque. Therefore, in our two-stream model, we could place the surface either above or below the long wave photosphere and replicate a physical situation on Earth. Our choice here to place the surface below the long wave photosphere is made in order to reproduce Earth's observed temperature-pressure profile (and mimicking a simple greenhouse model), which is the more important governing factor in the circulation of the atmosphere.
\subsection{Validation Model Results}
\begin{figure*}
    \includegraphics[width=0.325\textwidth]{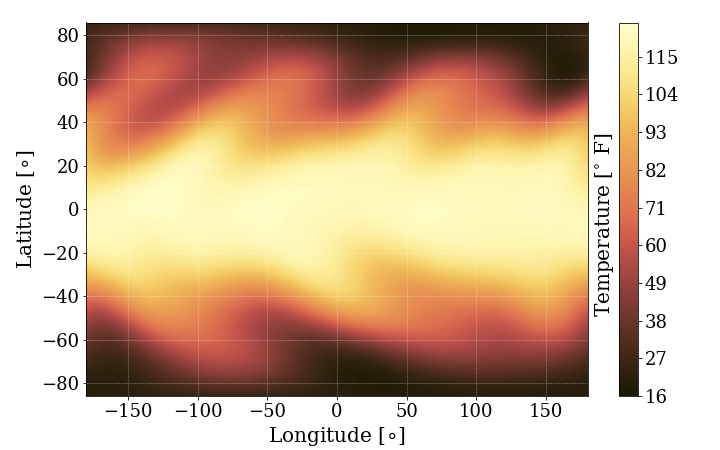}
    \includegraphics[width=0.325\textwidth]{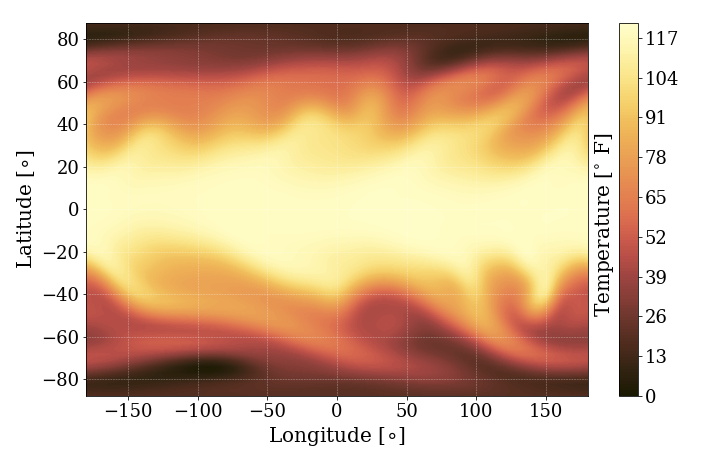}
    \includegraphics[width=0.325\textwidth]{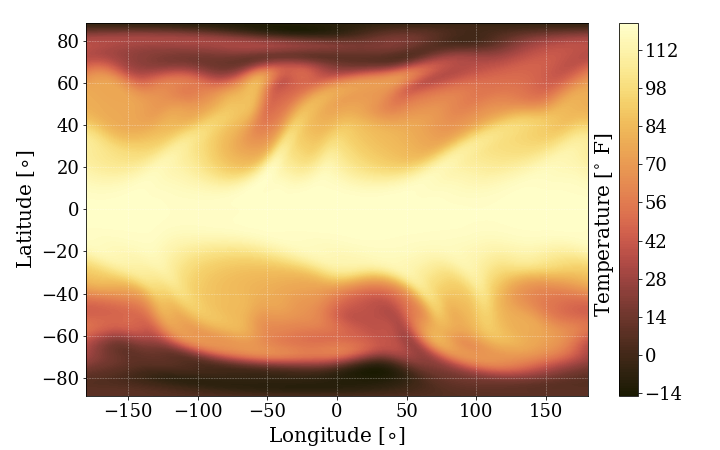}
    \includegraphics[width=0.325\textwidth]{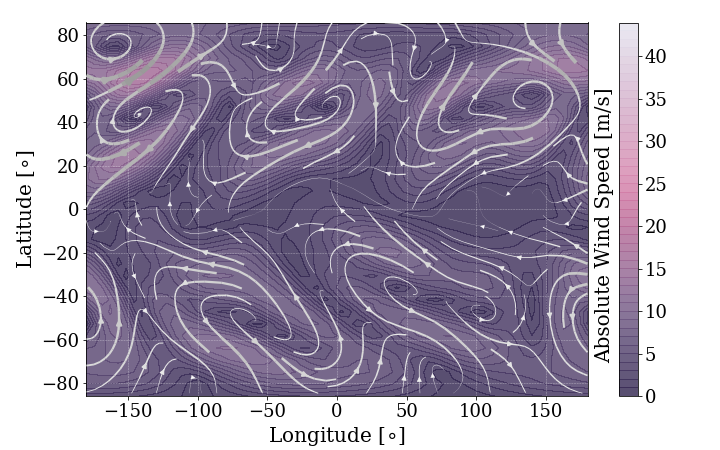}
    \includegraphics[width=0.325\textwidth]{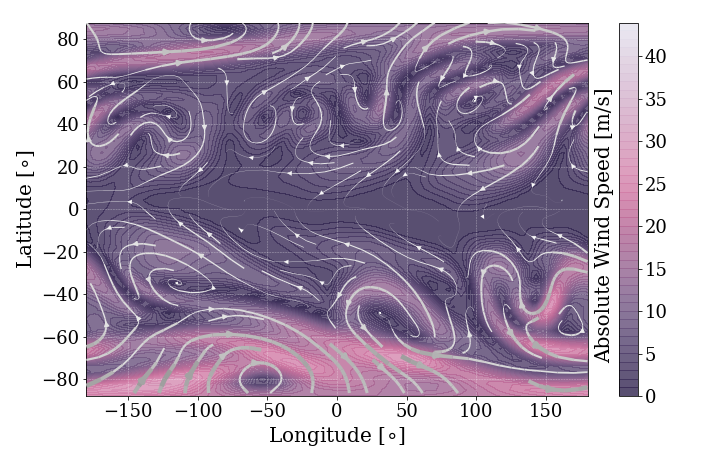}
    \includegraphics[width=0.325\textwidth]{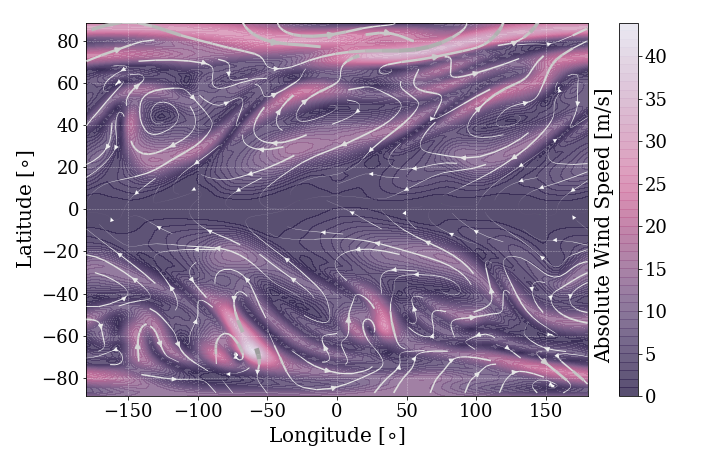}
    \caption{The top panels show temperature contours at the lowest atmospheric level (directly above the surface) for a single snapshot in time. The bottoms panels show a stream and contour plot of the winds on the same scale. \textbf{Left:} low resolution (T21) run, \textbf{Middle:} medium resolution (T42) run, \textbf{Right:} high resolution (T63) run. Notice that the relative scale of the temperatures is not dependent on model resolution, as expected. However, comparing the T21 and T42 runs suggests that the T21 run is not at a high enough resolution to fully resolve the baroclinic instabilities due to the surface.} \label{Surf:earth_temps}
\end{figure*}

\begin{figure}
	\plotone{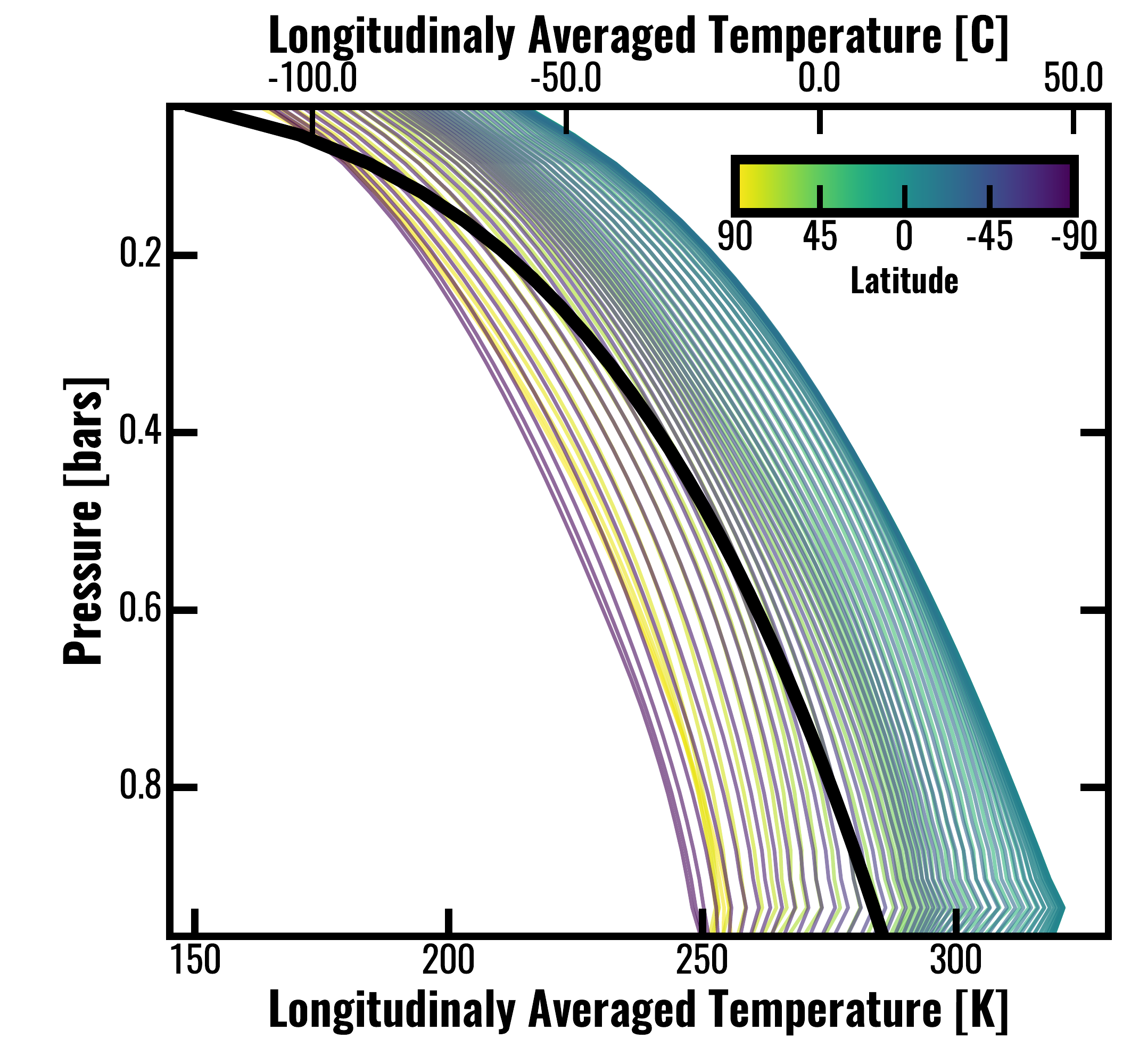}
	\caption{A comparison of our T63 resolution longitudinally averaged temperature-pressure profiles and a parameterized Earth troposphere model (black line, equation \ref{surf:earth_tp_param}).}
	\label{Surf:earth_tp_comp}
\end{figure}

We present circulation patterns from all three resolutions in Figure \ref{Surf:earth_temps}. After comparing our three resolutions, we find that while large-scale instabilities are present, the T21 resolution is not high enough to sufficiently reproduce the resulting circulation effects due to large and small scale baroclinic instabilities in the lower atmosphere. Figure \ref{Surf:earth_tp_comp} shows our longitudinally averaged temperature profiles from the T63 earth compared to a parameterized \citep{NEO} Earth temperature-pressure profile for the troposphere given as 
\begin{equation} \label{surf:earth_tp_param}
    p [KPa] = 101.29\times\left[\frac{T [K]}{288.08}\right]^{5.256}
\end{equation}
We see a deviation at low pressures, but attribute this to our model not extending past the tropopause region, while the parameterized model is defined to match with stratosphere temperatures. Overall, we find that our Earth model does a good job of replicating real-life conditions, even with its two stream double-grey radiative scheme and uniform surface slab assumptions. These results give us confidence that our updated RM12 GCM is sufficiently capturing the relevant surface-atmosphere interactions for terrestrial planets.

%
%
\section{Transition Planet} \label{Surf:TransitionPlanet}
\begin{figure}
	\plotone{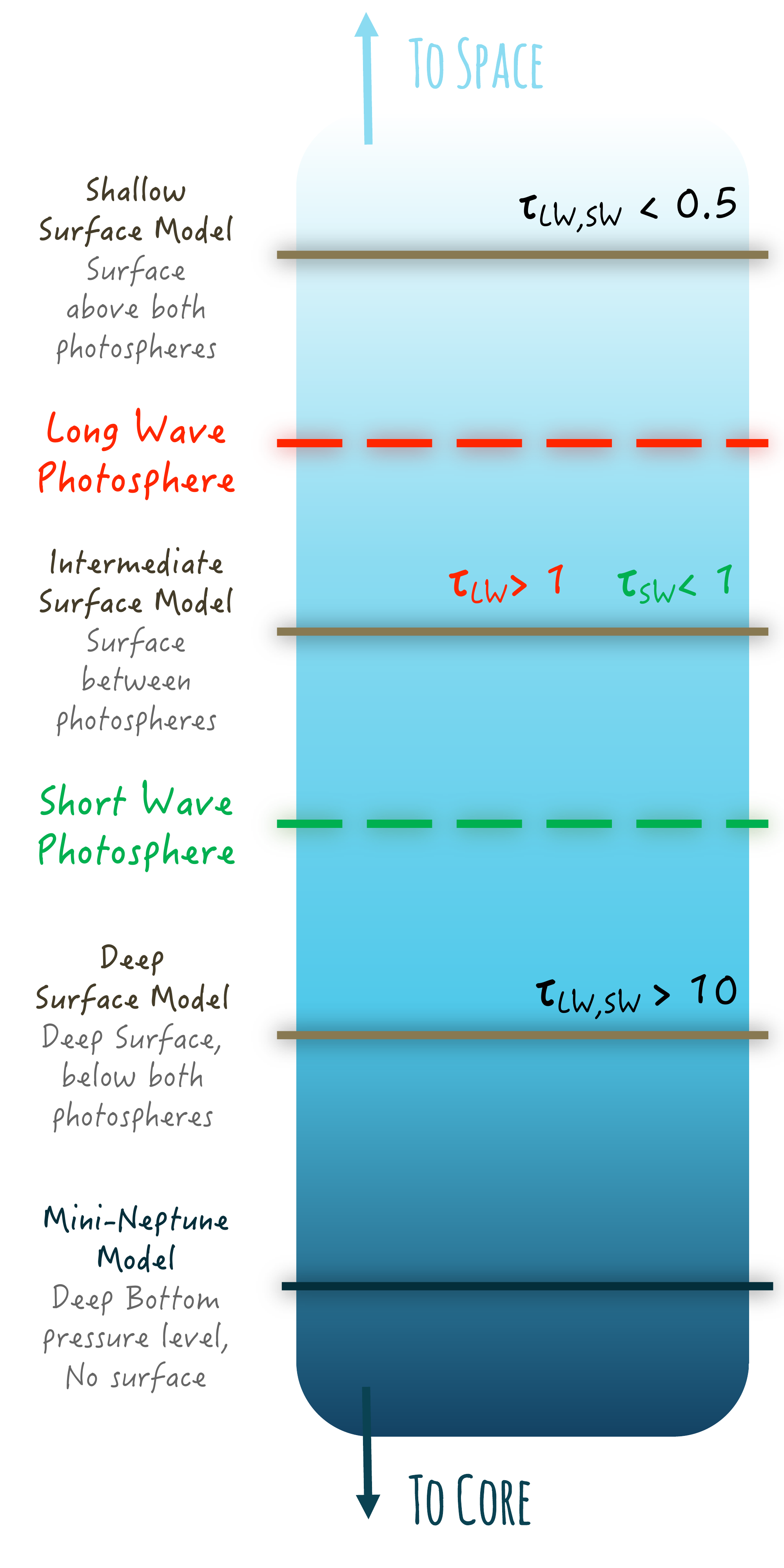}
	\caption{We define our four classes of models as shown. From bottom to top; the thickest atmosphere is our \textbf{Mini-Neptune Model}, with no surface included in the modeling; the \textbf{Deep Surface Model} has a surface placed at higher pressures than both the short wave and long wave photospheres; the \textbf{Intermediate Surface Model} has a surface placed between both photospheres; and the \textbf{Shallow Surface Model} has a surface above both photospheres.}
	\label{Surf:fig:schematic}
\end{figure}

We select a hypothetical planet at the transitional radius of 1.5 R$_{\oplus}$. For known exoplanets near the transition radius, error bars on measurements of their masses and radii and the breadth of composition curves that can match a given mass and radius result in large uncertainties in their Hydrogen-Helium mass fractions, placing them anywhere from a terrestrial planet with a thin atmosphere, to a gaseous planet with a thick atmosphere \citep{LopezFortney2014}. Therefore, we are justified in modeling this hypothetical 1.5 R$_{\oplus}$ transition planet as both terrestrial and gaseous with a wide range of surface pressures. We define four classes of transition planets, placing the surface at differing places in the atmosphere relative to the photospheres. Figure \ref{Surf:fig:schematic} shows a representation of these four classes of models. This replicates a range of Super-Earth to Mini-Neptune conditions. 
\subsection{Planetary Properties}
Our base planet has a radius of 1.5 R$_{\oplus}$ and a mass of 5 M$_{\oplus}$. We choose a Hydrogen/Helium dominated atmosphere with a similar metallicity to Neptune for consistency across all 4 of our base models. We select a long orbital period so that we may safely assume the planet is not synchronously rotating, therefore it is placed on a 100 day orbit around a solar-like star, with a rotation rate of 20 hours. With this selection, we are safe to use the same diurnally averaged heating scheme that is used for our Earth validation runs. The internal heat flux is set to 0.4 W m$^{-2}$. We select a resolution of T42 for all transition planets following estimates from Equations \ref{Surf:eqtn:rrd}-\ref{Surf:eqtn:ptp} of the mid-latitude Rossby Radius of Deformation, and from scaling off our Earth results. Table \ref{Surf:table:fidparam} summarizes the input parameters for our base models.

As in \cite{MR&F10,Miguel2014}, and following our Earth models, we set $\gamma$=0.7, where $\gamma$=$\kappa_{sw}/\kappa_{lw}$. For a predominately Hydrogen/Helium atmosphere, these authors set $\kappa_{lw}$=10$^{-2}$ cm$^{2}$ g$^{-1}$ giving us $\kappa_{sw}$=7$\times$10$^{-4}$ cm$^{2}$ g$^{-1}$. These values are similar to Earth, but the difference reflects that we choose to maintain a constant atmospheric composition of slightly higher than solar-metallicity across our transition regime, rather than an Earth-like atmosphere. From these values, our two photospheres are P$_{Photo,LW}=0.218$ bar and P$_{Photo,SW}=3.114$ bar. 

The difference between our 4 classes of models is primarily the surface pressure, which results in the atmospheric photospheres being at different locations relative to the solid surface and contributes to how the surface influences the circulation, particularly at the levels of the atmosphere we are sensitive to in observations. Table \ref{Surf:table:fidparam} lists the surface pressures (or bottom boundary for our Mini-Neptune model) for our Mini-Neptune, deep surface, intermediate surface, and shallow surface base models respectively; while Figure \ref{Surf:fig:schematic} shows a schematic representation of these classes of models. To maintain relative pressure resolution, the number of atmospheric levels is varied between models. We find that because of the effects of Rayleigh friction, vertical resolution becomes an important consideration for stability of the models. Our shallow surface and intermediate surface models are run with linear pressure levels due to their relatively thin atmospheres, with 10 and 40 levels, respectively. Our deep surface model and Mini-Neptune model are run with log pressure levels where the number of levels is calculated to maintain approximately the same number of levels above 2 bar as in the intermediate surface model. The result is 50 and 55 levels covering 3 orders of magnitude in pressure space, respectively. 
\begin{deluxetable}{lr} 
\tablecolumns{2}
\tablecaption{GCM Parameters for set of 4 Base Models \label{Surf:table:fidparam}}
\tablewidth{\textwidth}
\tablehead{
	\colhead{Parameter}		&
	\colhead{Value Adopted}			           	}
\startdata
            g (surface gravity)                  &       2180 cm s$^{-2}$                            \\
            R (specific gas constant)                   &       3779 J kg$^{-1}$ K$^{-1}$                   \\
            R$_p$ (planet radius)       &       9.56$\times$10$^{8}$ cm                     \\
            $\Omega$ (planet rotation rate)           &       8.73$\times$10$^{-5}$ radians s$^{-1}$      \\
            P$_0$ (surface pressure)              &       50 bar, 10 bar, 2 bar, 0.1 bar                 \\
            Vertical Levels               &       55, 50, 40, 10                  \\
            Pressure Orders of Magnitude &       3, 3, 0, 0                      \\
            F$_{int}$ (internal heat)          &       0.40 W m$^{-2}$                            \\
            F$_{irr}$ (stellar irradiation)          &       7680 W m$^{-2}$                             \\
            A (top of atmosphere albedo)                &       0.3                                         \\
            $\kappa_{sw}$ (visible absorption coefficient)       &       7.00$\times$10$^{-3}$ cm$^{2}$ g$^{-1}$     \\
            $\kappa_{lw}$ (thermal absorption coefficient      &       1.00$\times$10$^{-2}$ cm$^{2}$ g$^{-1}$      \\
            P$_{Photo,LW}$ (thermal photosphere)       &       0.218 bar    \\
            P$_{Photo,SW}$ (visible photosphere)      &       3.114 bar     
\enddata
    \tablecomments{Parameters for our set of base models. In rows with one value given, it is held constant across all four base models. In rows with four options listed is our input parameter for our Mini-Neptune, deep surface, intermediate surface, and shallow surface models, respectively.}
\end{deluxetable}
\subsection{Model Iterations}
To further explore the effects of a surface, we run iterations on our model classes by varying the surface pressure around the base model. For our deep surface model we run a total of 15 variations on the surface pressure ranging from 4 bar to 15 bar. If the surface pressure is less than 9.5 bar we run the model with 40 levels opposed to the standard 50 for computational speed. For the intermediate surface model we run a total of 13 variations on the surface pressure ranging from 0.25 bar to 3.0 bar. If the surface pressure is less than 2.0 bar the number of levels is computed such that the relative pressure resolution in constant between runs, with a minimum of 10 levels. Finally, for our shallow surface model, we run a total of 4 variations on surface pressure including the base model ranging from 0.01 bars to 0.20 bars, all with 10 levels. There are fewer variations for the shallow surface model due to numerical stability limitations for such a thin, low mass atmosphere. 
\begin{figure*}
    \centering
    \includegraphics[width=0.325\textwidth]{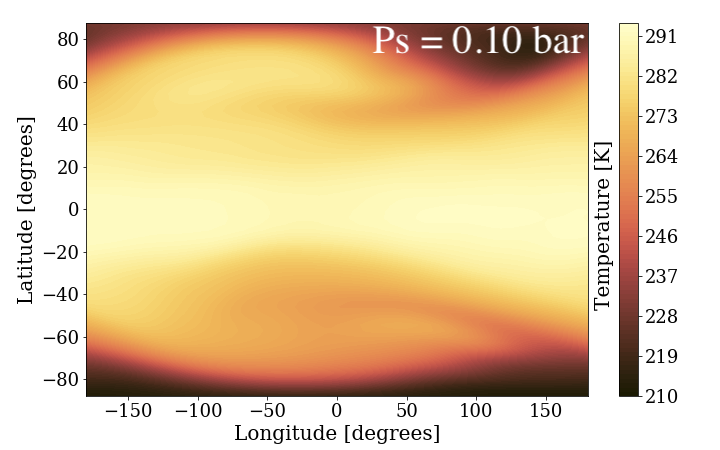}
    \includegraphics[width=0.325\textwidth]{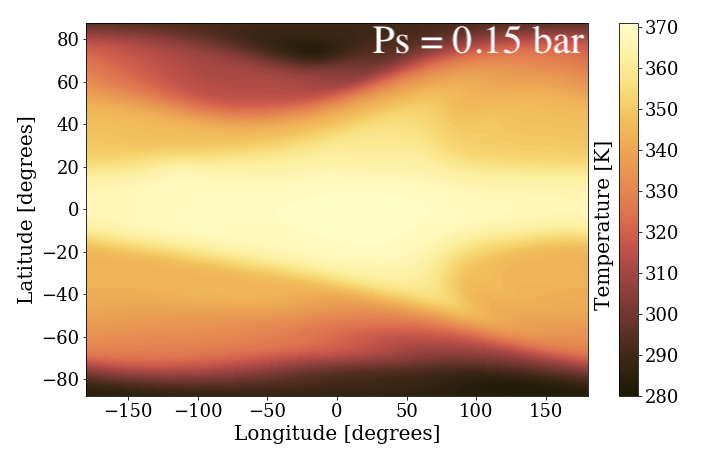}
    \includegraphics[width=0.325\textwidth]{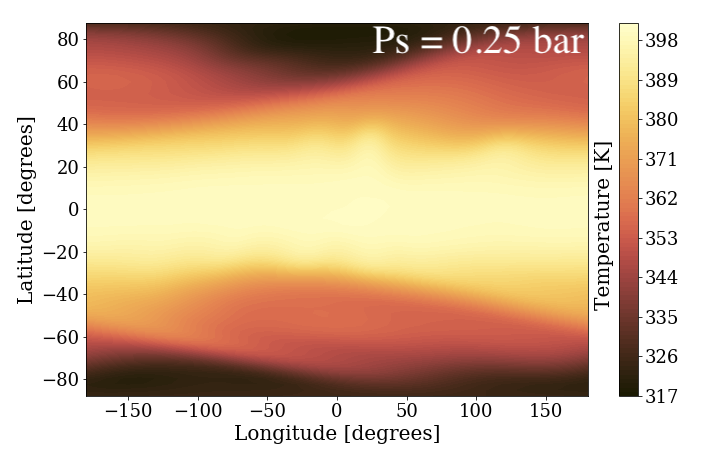}
    \includegraphics[width=0.325\textwidth]{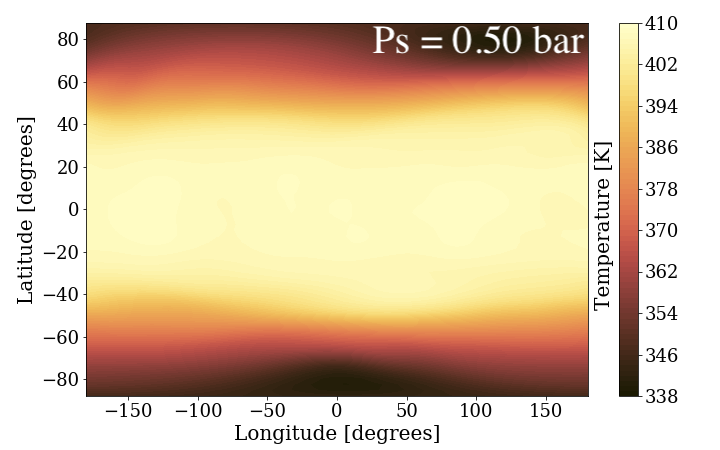}
    \includegraphics[width=0.325\textwidth]{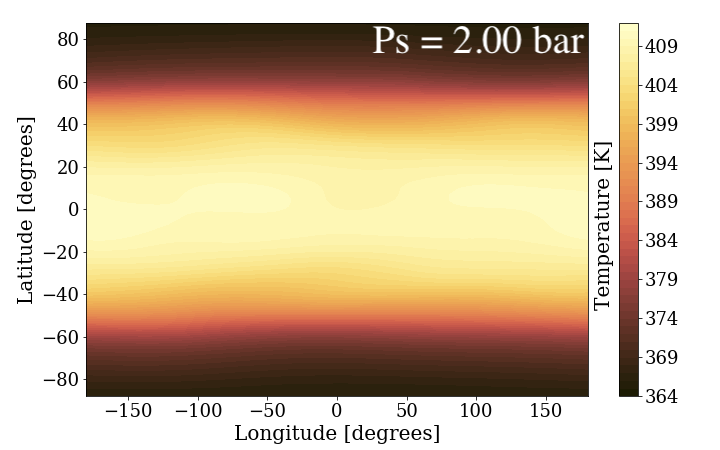}
    \includegraphics[width=0.325\textwidth]{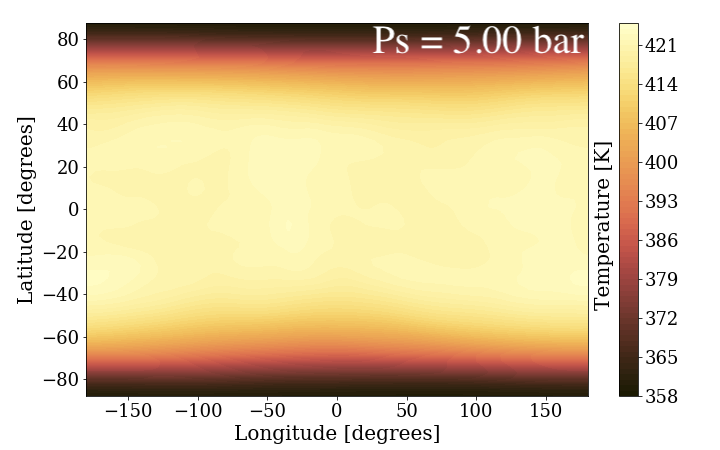}
    \caption{Long wave photospheric temperature maps for 6 of our model iterations (at 0.2 bar). For each panel the surface pressure is listed. Note the transition away from a baroclinic instability dominated flow once the surface moves below the short wave photosphere (at 3 bar) and no longer strongly affects the detectable levels in the atmosphere.}
    \label{Surf:model_temps}
\end{figure*}
\par Figure \ref{Surf:model_temps} shows temperature maps from the long wave photosphere for a selection of our model iterations. As listed in Table \ref{Surf:table:fidparam}, the long wave photosphere occurs at 0.218 bar. We see that as the surface moves deeper than this in the atmosphere, the effects of it on the overall atmospheric flow in the detectable levels of the atmosphere decrease, with the strength of the baroclinic instabilities dropping off significantly past a surface pressure of 0.25 bar. This is to be expected, and is a key feature that we hope to be able to distinguish observationally. Because baroclinic instabilities are more efficient at heat transport, atmospheres above a surface should have a lower equator-to-pole temperature (or emitted flux) difference. As the atmosphere becomes thicker, it becomes more difficult to distinguish the planet from a mini-Neptune. 
\section{Results: Transition Planet} \label{Surf:TPResults}
As discussed in the introduction, when considering observational implications, we are primarily interested in the various latitudinal dependencies of long wave emission and the reflected short wave light. Because of the position of the long wave photosphere, the shallow surface model and the iterations upon it are the only models where we directly see emission from the surface instead of the atmosphere itself. For all other models, we see emission from and above an atmospheric level corresponding to the long wave photosphere. For the shallow and intermediate surface models a significant fraction of the incoming stellar radiation reaches the surface and is reflected back into space. For the deep surface model, the surface is below the short wave photosphere and, aside from the top of the atmosphere reflection, no short wave radiation is reflected back to space, approaching the conditions of our Mini-Neptune model.

\begin{figure}
    \epsscale{1.3}
    \plotone{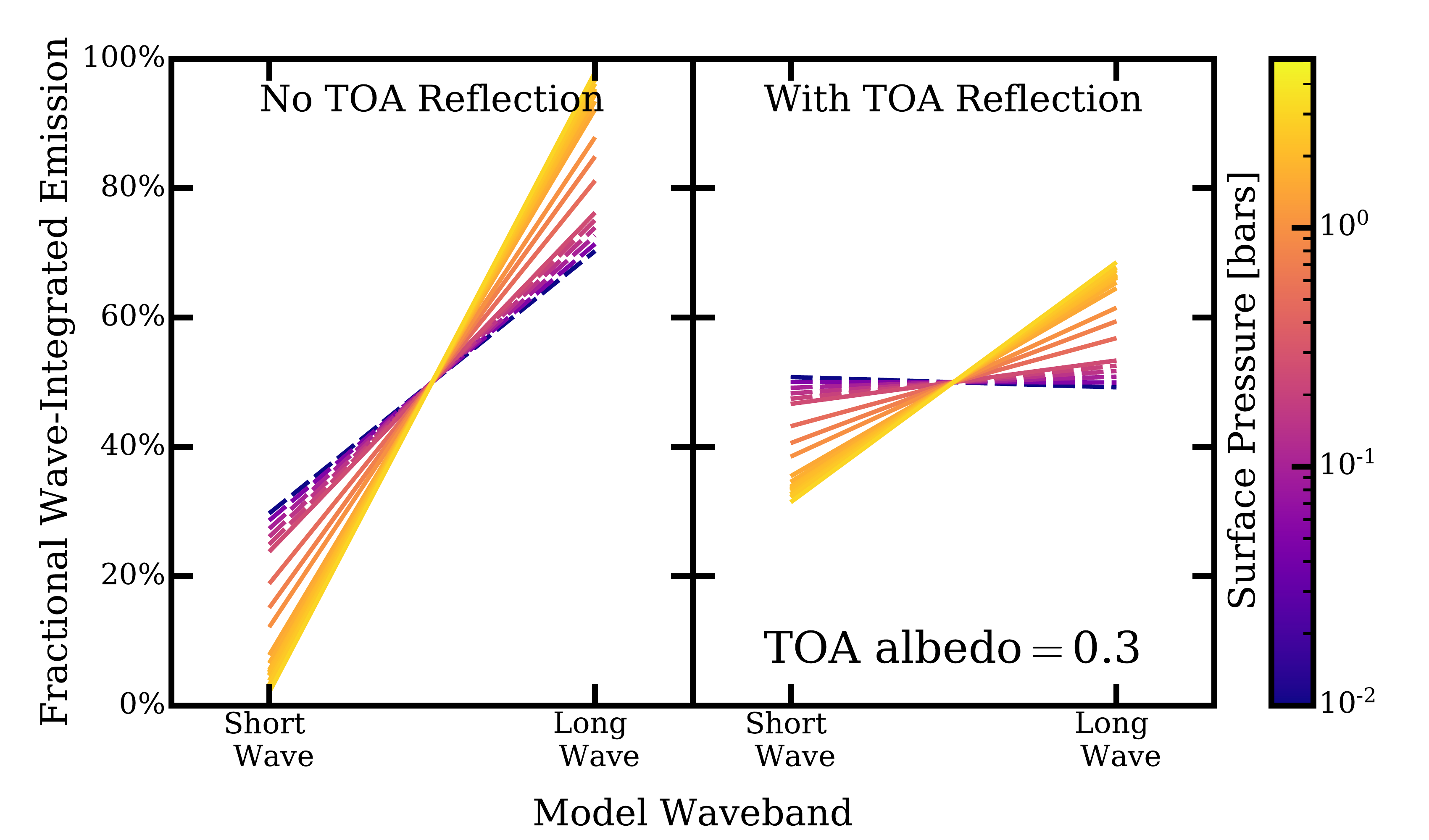}
	\caption{In both panels, the shallow surface models are shown by dashed lines and intermediate surface models are shown by solid lines. \textbf{Left:} Long wave emission and short wave surface reflection \textbf{Right:} The same, but taking into account the top of the atmosphere reflected short wave flux.}
	\label{Surf:fig:spec}
\end{figure}

\subsection{Band Integrated Emission and Top of the Atmosphere Albedos}
Observationally, the short wave surface reflection is combined with any top-of-the-atmosphere (TOA) reflection and cannot be independently determined without a-priori knowing the TOA albedo in our two-band model. There is a degeneracy between an atmosphere with a low TOA albedo and thin atmosphere (no TOA reflection, but a significant amount reflected from the surface) and an atmosphere with a high TOA albedo and a thick atmosphere (significant TOA reflection, but no surface reflection). Figure \ref{Surf:fig:spec} shows a dual band representation of the relative long wave emission and short wave surface + TOA reflection contributions to the planet's flux from our model runs. As expected, including TOA reflection results in a smaller difference between the long wave and short wave, and by comparing the shallow surface models without TOA reflection to the intermediate surface models with TOA reflection, one can see the discussed degeneracy. Because of this degeneracy, our two-band model suggests that simply measuring the disk integrated relative reflected short wave to emitted long wave light is not a robust way to determine the presence of a surface unless the TOA albedo can be determined in some other way. 

\cite{Demory2014} calculate geometric albedos for Super-Earths in the Kepler sample assuming there is no reflected short wave from the surface, which is a good assumption for most of the planets in the sample due to their radii being large enough that they are more likely to be gaseous, mini-Neptunes. However, as several planets in the sample skirt the line of terrestrial vs. gaseous compositions and may have intermediate to shallow surfaces, the contribution from surface reflection cannot be ignored. Other work to measure the top of the atmosphere albedo focuses on large gaseous planets where all short wave light detected should be in the form of TOA reflection \citep[for example,][]{Anger2015,Bell2017,Mallonn2019}. More work needs to be done on disentangling the surface reflection before we can use spectra to determine the presence of a surface for transition regime planets. However, in the case of clouds and tidal locking, as shown by \cite{Mansfield2019} and \cite{Koll2019}, the high albedo of the clouds allows the inference of an atmosphere, although the presence of an atmosphere on a 1.5 R$_{\oplus}$ planet alone does not differentiate it between a terrestrial or gaseous composition.

While two-band eclipses themselves are uninformative as to the presence of a surface from our modeling choices, it is important to discus our ability to detect such planets in eclipse since the mapping of latitudinal variations as discussed in the next section is without hope if the eclipse itself is beyond our limits. In this paper, we choose to place our planets around a Sun-like star to study the effects of a surface on Earth-similar worlds. Naturally, the detection of even a transit of a Earth-similar planet around a solar like star is difficult with transit depths of $\sim$ 0.01\%. For the temperate planet of 1.5 R$_{\oplus}$ in our models, the secondary eclipse depths are of order 0.001 ppm (1 part per trillion), clearly beyond the possibility of detection with any upcoming missions.  

Because we cannot detect the secondary eclipse of the modeled planets around a sun-like star with upcoming mission specifications, we instead scale our model results to a planet around an `average' M-dwarf receiving the same total amount of top-of-the atmosphere irradiation. While the exact contribution of short wave and long wave components to the instellation will be different due to the stellar radiation shifting to longer wavelengths, our double-grey model does not take this into account and applies all instellation as short wave. Therefore our scaling of eclipse depths is sufficient to capture the trend from a surface to a gaseous planet as our model inputs that control the heating would be exactly the same for a planet around this `average' M-dwarf. 

\begin{figure}
    \includegraphics[width=0.45\textwidth]{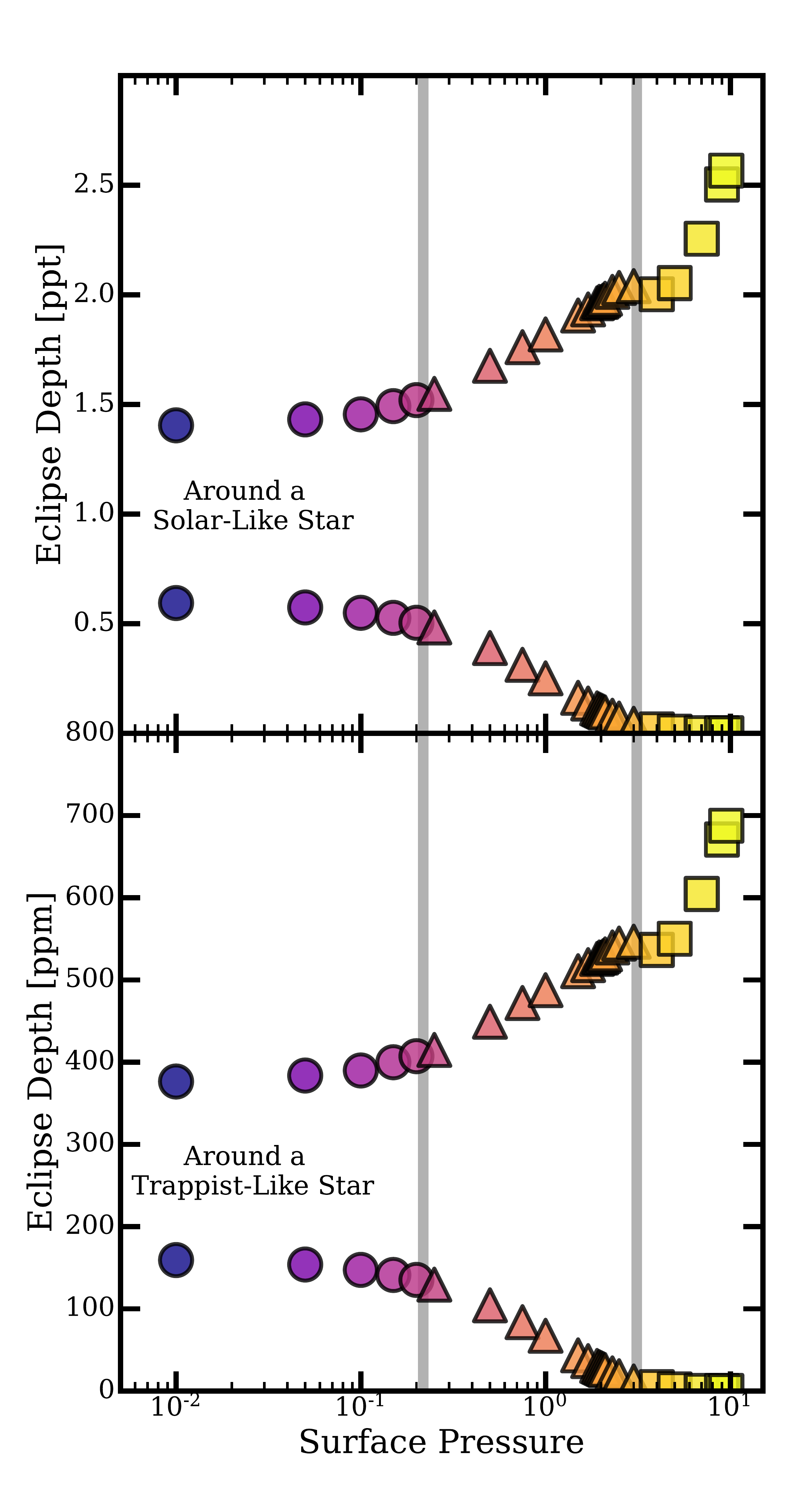}
	\caption{In both panels, the long wave (top curve) and short wave (bottom curve) calculated eclipse depths for our suite of models. The symbols represent the different classes of models with circles, triangles, and squares corresponding to our shallow surface, intermediate surface, and deep surface models, respectively. \textbf{Top}: around the modeled sun-like host star in parts-per-trillion and \textbf{bottom}: scaled to a TRAPPIST-1 like star in parts-per-million. The grey vertical lines correspond the the long wave and short wave photospheres.}
	\label{Surf:fig:edepth}
\end{figure}

Under these conditions, the eclipse depth itself becomes easily detectable at $\sim$100s ppm. Figure \ref{Surf:fig:edepth} shows the calculated long wave (top curves) and short wave (bottom curves) eclipse depths for our various model iterations around the modeled sun-like host star and scaled to a TRAPPIST-1 like star. When discussing the observability of the latitudinal differences, we maintain this scaling to a TRAPPIST-1 like star.

As Figure \ref{Surf:fig:edepth} shows, the long wave eclipses become deeper as the surface moves deeper in the atmosphere, corresponding to more of the incoming short wave radiation being absorbed and emitted in the long wave channel. The short wave eclipses become shallower and eventually reach a depth of 0 ppm as all of the incoming short wave radiation is absorbed throughout the atmosphere and none is left to be reflected off the surface. Note that this figure does not include any top-of-the-atmosphere reflection in the short wave eclipse depths, but in our modeling scheme this would simply be a constant offset.
\subsection{Zonal Averaged Emission and Reflection}
\begin{figure*}
    \epsscale{1.2}
    \plotone{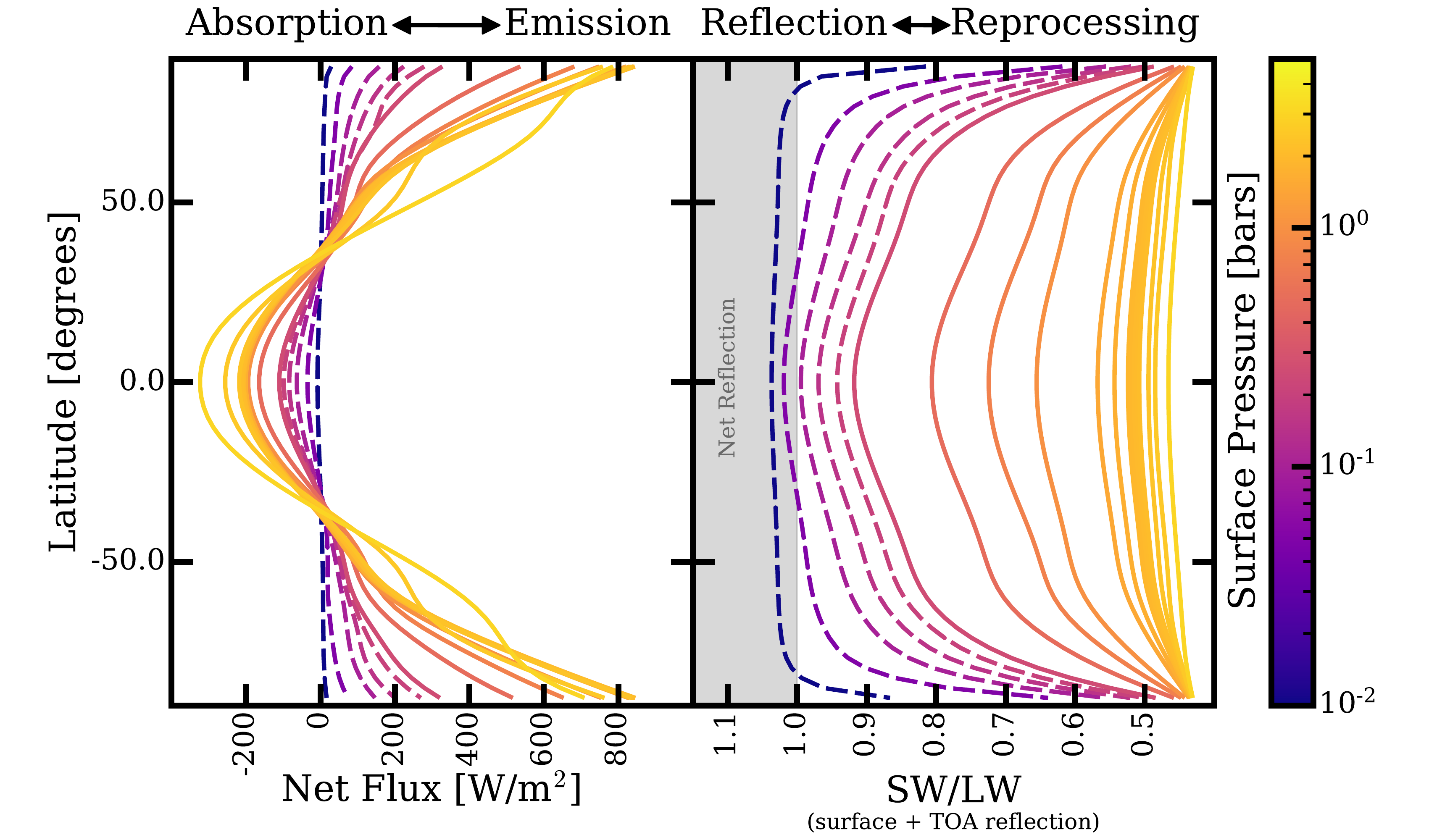}
	\caption{In both panels, the shallow and intermediate surface models are shown by dashed and solid lines, respectively. \textbf{Left:} Net flux at the top of the atmosphere, where positive values indicate emission to space. \textbf{Right:} Ratio of the total reflected to emitted light; the shaded region represents where reflection is greater than emission.}
	\label{Surf:fig:LWE}
\end{figure*}

While our two-band model shows degeneracies in band-integrated light, we have already shown in Figure \ref{Surf:model_temps} that there are latitudinal differences between the various models in the detectable levels of the atmosphere. To that means, Figure \ref{Surf:fig:LWE} presents the zonal (longitudinal) average net fluxes (left) and relative short-wave to long-wave emission/reflection (right) for all shallow and intermediate surface models, averaged over their last modeled orbit. The two panels share a common colorbar with dashed lines representing those with surfaces above the long wave photosphere and solid lines representing those with surface below the long wave photosphere

As shown in the left panel, for the shallowest atmosphere (0.01 bars, dashed purple line), the net flux is approximately constant, meaning that heat is primarily emitted (reflected) back to space at the latitudes it was received, with little heat transport or reprocessing in the atmosphere. As we move to deeper surfaces (higher surface pressures, solid yellow lines), we find that the relative amount of net absorption at the equator compared to net emission near the poles becomes more distinct, corresponding to more efficient atmospheric heat transport away from the equator, as one would expect. Further, for surfaces near the long wave photosphere where we more directly can observe the influence of the surface, there is a flattening off at mid latitudes representing heat being moved more efficiently away from these latitudes towards the poles as a result of heat transport through eddy formation. As the surface moves deeper, this effect is less pronounced due to the effects of the surface mostly occurring below the levels from which we see emission. 

In the right panel of Figure \ref{Surf:fig:LWE}, we see that as the surface moves deeper the total short wave surface+TOA reflection becomes less important in comparison to the long wave emission. As discussed, while the disk-integrated short-wave to long-wave ratio for a planet with a surface is degenerate with the contribution from a uniform TOA albedo (see Figure \ref{Surf:fig:spec}), the qualitative shape of this ratio as a function of latitude depends on the presence of a surface and is unaffected by the uniform TOA albedo. For models with shallow surfaces (above the long wave photosphere, dashed purple and pink lines), the total short-wave-reflection to long-wave-emission ratio is relatively constant with latitude until a sharp drop off near the poles, and the ratio becomes more dependent on latitude as the surface moves deep enough to allow for atmospheric heat redistribution (solid yellow and orange lines).

An extention of this discussion is the equator-to-pole emission difference. While not shown here, we find that this value as a function of model surface pressure is qualitatively similar to that of \cite{KS15} and \cite{Komacek2019} wherein the measured equator-to-pole difference decreases with increasing surface pressure, as atmospheric heat transport becomes more efficient. However, neither of these works explicitly presents zonally averaged comparisons of planets in this transition regime and discuss the observational implications of a surface. We further fill out the surface pressure dimension with detailed comparisons of various classes of models and how heat transport changes within this regime. 

The curves presented in Figure \ref{Surf:fig:LWE} and the equator-to-pole emitted light difference is an observable quantity with secondary eclipse mapping as discussed in the introduction to this paper. Typical errors on measurements with Spitzer IRAC are too high to resolve temperature structures in a small number of orbits for most planets, however the expected noise floor of JWST is 15-30 ppm which is sufficient to resolve the eclipse depth of a temperate Super-Earth in 25 transits \citep{Beichman2014}, however the ability to resolve temperatures through eclipse mapping is more difficult than simply detecting the eclipse itself. 

K2-18b has been a planet of much discussion lately due to the recent detection of water-vapour in it's atmosphere \citep{Benneke2019,Tsiaras2019}. Although it is the smallest planet with such a measurement to-date, at 2.6 R$_{\oplus}$, it is well outside the radius regime of the transition from Super-Earths and Mini-Neptunes, and any surface, if one existed, would be below such a thick atmosphere that no reflection or emission off of a surface would be detectable. To demonstrate this, \cite{Madhusudhan2020} calculate possible compositions to explain to mass/radius of K2-18b and suggest several possibilities, including (1) a rocky core with 5\% mass in the H/He envelope, yielding an atmospheric pressure of $\sim$ 10$^6$ bar, and (2) a water world with a deep ocean and a relatively small envelope of H/He (0.006\% by mass) corresponding to a pressure of 100s of bars at the transition between the two components. Of particular interest to this planet are therefore our deep surface models, which have an atmospheric pressure of 10s of bars at the surface boundary, and our  mini-neptune model which includes no surface and a bottom pressure boundary of 50 bars. For the atmospheric composition we have selected, the thermal and visible photospheres of 0.2 and 3.1 bars, respectively, are high enough in the atmosphere relative to the deepest pressures in these cases that no incoming short wave radiation reaches the bottom boundaries, and therefore there is no surface reflection component to be detected in the case of a surface. The emission seen from these models is from a level so detached from any surface, that as shown in Figure \ref{Surf:model_temps}, we see no baroclinic instabilities forming at these levels, and therefore no enhanced equator-to-pole heat transport. There is no realistic composition for K2-18b which would place it in our shallow or intermediate surface model cases where a surface impacts the detectable regions of the atmosphere in a measurable way.

\subsection{Secondary Eclipse Mapping to Detect a Surface}
To determine the scale of the latitudinal signal we hope to measure with eclipse mapping we use the python package spiderman \citep{Louden2017} to generate secondary eclipses from our GCM output long wave and short wave emission maps.  These eclipses are compared to those that would exist for a uniform sphere with the same absolute eclipse depth. The difference between these curves is in the shapes of ingress and egress which contain information about the emission map from the planet. The scale of the differences between the model eclipse and a uniform sphere eclipse quantifies our ability to detect latitudinal variations from our suite of 1.5 R$_{Earth}$ planets. 

\begin{figure}
    \epsscale{1.2}
    \plotone{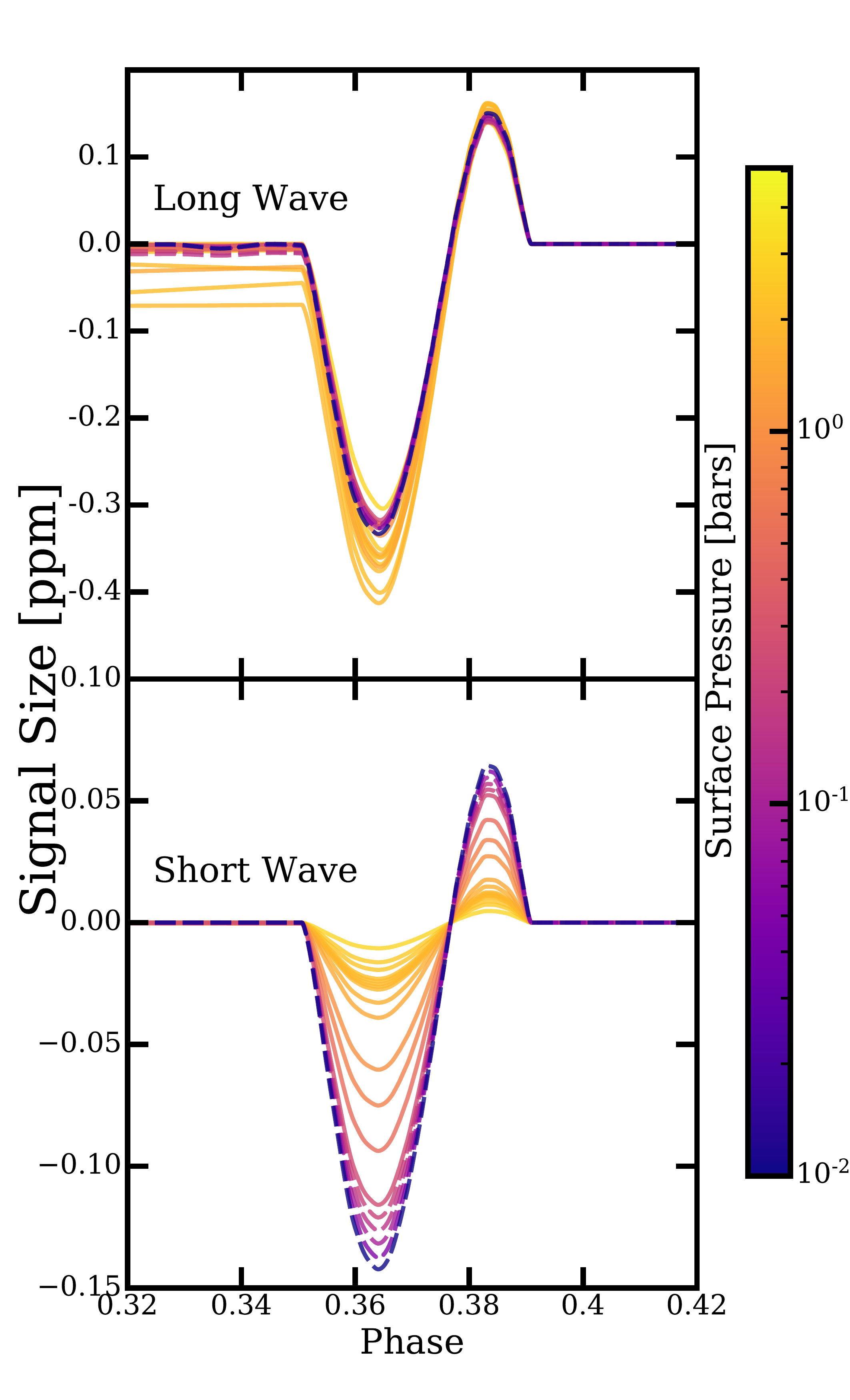}
	\caption{The difference between the ingress curve (for secondary eclipse) of a uniform disk and the spatial maps from our suite of models. In both panels, the shallow and intermediate surface models are shown by dashed and solid lines, respectively. \textbf{Top:} Long wave eclipse mapping signal \textbf{Bottom:} Short wave eclipse mapping signal. For all models a non-uniform flux pattern would be seen in thermal (long wave) emission, whereas only in the shallow atmosphere cases do we expect latitudinal variation in the reflected (short wave) flux.}
	\label{Surf:fig:EM}
\end{figure}

In Figure \ref{Surf:fig:EM} we present eclipse depth signals for the various surface pressure models scaled to an M-dwarf and zoomed in on eclipse ingress. 
In the top panel we show the long wave signal and in the bottom panel we show the short wave signal. We find a larger variation between the models for short wave eclipses, with the effect going to zero as the short wave reflection off the surface approaches zero and the planet appears to become a uniform sphere when assuming a uniform TOA albedo. 
This suggests that if an eclipse map at short wavelengths shows a latitudinal dependence, there is likely reflection off of a surface contributing to the observed features. However, we note that the scale of these signals is at best 0.1 ppm, below the noise floors of any planned missions. The long wave eclipse map signal differences are less obvious because the shape is unchanging with surface pressure. We see that as the surface moves deeper in the atmosphere the signal size increases, suggesting a planet more different from a uniform sphere, or a larger equator to pole temperature difference. While the maximum eclipse map signal is similar to that of the short wave signal, the difference between the models is even smaller. 

Together, the maps that can be generated from a short wave and long wave eclipse would allow a determination of the ratio between the zonally averaged short wave and long wave light as shown in Figure \ref{Surf:fig:LWE} which would place constraints on the presence of a surface.

%
\section{Conclusions} \label{Surf:Conclusions}
In this paper we have explored the observational effects a surface imparts on an atmosphere for non-tidally synchronized planets with radii intermediate between Earth and Neptune (i.e., the transition regime between terrestrial and gaseous). Because there is a degeneracy in global composition from a measurement of density alone, and because density is hard to measure precisely anyway, we sought to determine an alternative method to classify these transition regime planets. In particular we have explored
\begin{itemize}
    \item Disk integrated emission and reflection in our long wave/short wave band passes -- We find that our two band model produces a degeneracy between top of the atmosphere (TOA) reflection and the thickness of the atmosphere, with a thick atmosphere and a high TOA albedo looking identical to a thin atmosphere with a low TOA albedo. We expect that this degeneracy would not exist for a more finely sampled wavelength space as cloud/TOA reflection should impart a different signal than surface reflection. 
    \item Zonally averaged emission and reflection in our long wave/short wave band passes -- We find that the location of the surface relative to the long wave photosphere plays an important roll in the equator-to-pole flux differences as expected and previously reported \citep{KS15,Komacek2019}. A key feature being that for the shallowest atmospheres modeled the net incoming/outgoing flux is approximately constant with latitude with a large equator-to-pole temperature or emitted flux difference. We see a quick drop off in this equator-to-pole difference as the atmosphere becomes thicker and can more efficiently transport heat away from the equator, with those planets with surfaces near the long wave photosphere showing strong heat transport at mid latitudes due to baroclinic eddy formation. If the eclipse can be highly resolved enough to be mapped, the shape of the short wave to long wave ratio will be a key factor in determining the presence of a surface.
    \item The detectability of a surface using secondary eclipse observations -- While the secondary eclipses alone are easily detectable for a planet of this temperature around an M-dwarf, we have shown that disk integrated light in two bands is degenerate between a surface and TOA albedo. We have further shown that eclipse mapping can break this degeneracy and that a signature of a surface is embedded in these maps. However, this signal is small ($\sim$0.1 ppm around an M-dwarf) and is not detectable with any current or planned mission. 
\end{itemize}
We further clarify that the results presented here are applicable only to the specific modeling choices made, and that there are perhaps regimes of  parameter space where the latitudinal variations due to the surface produce a larger, more detectable signal. For example, studies of the eclipse mapping feature size for models with slab oceans as opposed to our solid surface, in combination with the moist effects such a modeling choice necessitates, is a possible future direction to further explore the observational differences between super-Earths and mini-Neptunes within the radius valley. Still, we expect this signal to be small and outside the reach of JWST or the next generation large mission concepts in development now. 

Disk integrated light, while inconclusive in our work, will likely be a powerful way to determine if a planet is rocky or gaseous by measuring eclipses over more than 2 bands. In particular, spectroscopic observations or simultaneous band passes will remain observationally cheap compared to full phase curve measurements to detect atmospheres. Further work on this topic is necessary to determine key signatures of surface vs. TOA and cloud reflection for non-clear atmospheres.

\acknowledgements{This research was supported by NASA Astrophysics Theory Program grant NNX17AG25G. The first author acknowledges Space Telescope Science Institute for temporarily hosting them during the completion of this article.}
%

\begin{thebibliography}{}
\expandafter\ifx\csname natexlab\endcsname\relax\def\natexlab#1{#1}\fi

\bibitem[{{Angerhausen} {et~al.}(2015){Angerhausen}, {DeLarme}, \&
  {Morse}}]{Anger2015}
{Angerhausen}, D., {DeLarme}, E., \& {Morse}, J.~A. 2015, \pasp, 127, 1113

\bibitem[{{Batalha}(2014)}]{Batalha2014}
{Batalha}, N.~M. 2014, Proceedings of the National Academy of Science, 111,
  12647

\bibitem[{{Beichman} {et~al.}(2014){Beichman}, {Benneke}, {Knutson}, {Smith},
  {Dressing}, {Latham}, {Deming}, {Lunine}, {Lagage}, {Sozzetti}, {Beichman},
  {Sing}, {Kempton}, {Ricker}, {Bean}, {Kreidberg}, {Bouwman}, {Crossfield},
  {Christiansen}, {Ciardi}, {Fortney}, {Albert}, {Doyon}, {Rieke}, {Rieke},
  {Clampin}, {Greenhouse}, {Goudfrooij}, {Hines}, {Keyes}, {Lee}, {McCullough},
  {Robberto}, {Stansberry}, {Valenti}, {Deroo}, {Mand ell}, {Ressler},
  {Shporer}, {Swain}, {Vasisht}, {Carey}, {Krick}, {Birkmann}, {Ferruit},
  {Giardino}, {Greene}, \& {Howell}}]{Beichman2014}
{Beichman}, C., {Benneke}, B., {Knutson}, H., {et~al.} 2014, arXiv e-prints,
  arXiv:1411.1754

\bibitem[{{Bell} {et~al.}(2017){Bell}, {Nikolov}, {Cowan}, {Barstow}, {Barman},
  {Crossfield}, {Gibson}, {Evans}, {Sing}, {Knutson}, {Kataria}, {Lothringer},
  {Benneke}, \& {Schwartz}}]{Bell2017}
{Bell}, T.~J., {Nikolov}, N., {Cowan}, N.~B., {et~al.} 2017, \apj, 847, L2

\bibitem[{{Belu} {et~al.}(2011){Belu}, {Selsis}, {Morales}, {Ribas}, {Cossou},
  \& {Rauer}}]{Belu2011}
{Belu}, A.~R., {Selsis}, F., {Morales}, J.~C., {et~al.} 2011, \aap, 525, A83

\bibitem[{{Benneke} {et~al.}(2019){Benneke}, {Wong}, {Piaulet}, {Knutson},
  {Lothringer}, {Morley}, {Crossfield}, {Gao}, {Greene}, {Dressing},
  {Dragomir}, {Howard}, {McCullough}, {Kempton}, {Fortney}, \&
  {Fraine}}]{Benneke2019}
{Benneke}, B., {Wong}, I., {Piaulet}, C., {et~al.} 2019, \apjl, 887, L14

\bibitem[{{Charbonneau} {et~al.}(2009){Charbonneau}, {Berta}, {Irwin}, {Burke},
  {Nutzman}, {Buchhave}, {Lovis}, {Bonfils}, {Latham}, {Udry}, {Murray-Clay},
  {Holman}, {Falco}, {Winn}, {Queloz}, {Pepe}, {Mayor}, {Delfosse}, \&
  {Forveille}}]{Charbonneau2009}
{Charbonneau}, D., {Berta}, Z.~K., {Irwin}, J., {et~al.} 2009, \nat, 462, 891

\bibitem[{{de F. Forster} {et~al.}(2000){de F. Forster}, {Blackburn}, {Glover},
  \& {Shine}}]{deForster2000}
{de F. Forster}, P.~M., {Blackburn}, M., {Glover}, R., \& {Shine}, K.~P. 2000,
  Climate Dynamics, 16, 833

\bibitem[{{Demory}(2014)}]{Demory2014}
{Demory}, B.-O. 2014, \apj, 789, L20

\bibitem[{{Flowers} {et~al.}(2019){Flowers}, {Brogi}, {Rauscher}, {Kempton}, \&
  {Chiavassa}}]{Flowers2018}
{Flowers}, E., {Brogi}, M., {Rauscher}, E., {Kempton}, E. M.~R., \&
  {Chiavassa}, A. 2019, \aj, 157, 209

\bibitem[{{Frierson} {et~al.}(2006){Frierson}, {Held}, \&
  {Zurita-Gotor}}]{Frierson2006}
{Frierson}, D.~M.~W., {Held}, I.~M., \& {Zurita-Gotor}, P. 2006, Journal of
  Atmospheric Sciences, 63, 2548

\bibitem[{{Fulton} {et~al.}(2017){Fulton}, {Petigura}, {Howard}, {Isaacson},
  {Marcy}, {Cargile}, {Hebb}, {Weiss}, {Johnson}, {Morton}, {Sinukoff},
  {Crossfield}, \& {Hirsch}}]{Fulton2017}
{Fulton}, B.~J., {Petigura}, E.~A., {Howard}, A.~W., {et~al.} 2017, \aj, 154,
  109

\bibitem[{{Guillot}(2010)}]{Guillot2010}
{Guillot}, T. 2010, \aap, 520, A27

\bibitem[{{Held} \& {Suarez}(1994)}]{HS94}
{Held}, I.~M., \& {Suarez}, M.~J. 1994, Bulletin of the American Meteorological
  Society, 75, 1825

\bibitem[{{Holton}(1992)}]{HoltonBook}
{Holton}, J.~R. 1992, {An introduction to dynamic meteorology}

\bibitem[{{Hoskins} \& {Simmons}(1975)}]{Hoskins1975}
{Hoskins}, B.~J., \& {Simmons}, A.~J. 1975, Quarterly Journal of the Royal
  Meteorological Society, 101, 637

\bibitem[{{Joshi} {et~al.}(1995){Joshi}, {Lewis}, {Read}, \&
  {Catling}}]{Joshi1995}
{Joshi}, M.~M., {Lewis}, S.~R., {Read}, P.~L., \& {Catling}, D.~C. 1995,
  Journal of Geophysical Research, 100, 5485

\bibitem[{{Kanodia} {et~al.}(2019){Kanodia}, {Wolfgang}, {Stefansson}, {Ning},
  \& {Mahadevan}}]{Kanodia2019}
{Kanodia}, S., {Wolfgang}, A., {Stefansson}, G.~K., {Ning}, B., \& {Mahadevan},
  S. 2019, arXiv e-prints, arXiv:1903.00042

\bibitem[{{Kaspi} \& {Showman}(2015)}]{KS15}
{Kaspi}, Y., \& {Showman}, A.~P. 2015, \apj, 804, 60

\bibitem[{{Kataria} {et~al.}(2016){Kataria}, {Sing}, {Lewis}, {Visscher},
  {Showman}, {Fortney}, \& {Marley}}]{Kataria2016}
{Kataria}, T., {Sing}, D.~K., {Lewis}, N.~K., {et~al.} 2016, \apj, 821, 9

\bibitem[{{Koll} \& {Abbot}(2016)}]{Koll2016}
{Koll}, D. D.~B., \& {Abbot}, D.~S. 2016, \apj, 825, 99

\bibitem[{{Koll} {et~al.}(2019){Koll}, {Malik}, {Mansfield}, {Kempton}, {Kite},
  {Abbot}, \& {Bean}}]{Koll2019}
{Koll}, D. D.~B., {Malik}, M., {Mansfield}, M., {et~al.} 2019, arXiv e-prints,
  arXiv:1907.13138

\bibitem[{{Komacek} \& {Abbot}(2019)}]{Komacek2019}
{Komacek}, T.~D., \& {Abbot}, D.~S. 2019, \apj, 871, 245

\bibitem[{{Komacek} {et~al.}(2019){Komacek}, {Jansen}, {Wolf}, \&
  {Abbot}}]{Komacek2019b}
{Komacek}, T.~D., {Jansen}, M.~F., {Wolf}, E.~T., \& {Abbot}, D.~S. 2019, \apj,
  883, 46

\bibitem[{{Komacek} \& {Showman}(2016)}]{Komacek2016}
{Komacek}, T.~D., \& {Showman}, A.~P. 2016, \apj, 821, 16

\bibitem[{{Kreidberg} {et~al.}(2014){Kreidberg}, {Bean}, {D{\'e}sert},
  {Benneke}, {Deming}, {Stevenson}, {Seager}, {Berta-Thompson}, {Seifahrt}, \&
  {Homeier}}]{Kreidberg2014}
{Kreidberg}, L., {Bean}, J.~L., {D{\'e}sert}, J.-M., {et~al.} 2014, \nat, 505,
  69

\bibitem[{{Kreidberg} {et~al.}(2019){Kreidberg}, {Koll}, {Morley}, {Hu},
  {Schaefer}, {Deming}, {Stevenson}, {Dittmann}, {Vanderburg}, {Berardo},
  {Guo}, {Stassun}, {Crossfield}, {Charbonneau}, {Latham}, {Loeb}, {Ricker},
  {Seager}, \& {Vand erspek}}]{Kreidberg2019}
{Kreidberg}, L., {Koll}, D. D.~B., {Morley}, C., {et~al.} 2019, \nat, 573, 87

\bibitem[{Krupka {et~al.}(1985)Krupka, Hemingway, Borie, \&
  Kerrick}]{Krupka1985}
Krupka, K., Hemingway, B., Borie, R., \& Kerrick, D. 1985, 70, 261

\bibitem[{{Lopez} \& {Fortney}(2014)}]{LopezFortney2014}
{Lopez}, E.~D., \& {Fortney}, J.~J. 2014, \apj, 792, 1

\bibitem[{{Louden} {et~al.}(2017){Louden}, {Wheatley}, {Irwin}, {Kirk}, \&
  {Skillen}}]{Louden2017}
{Louden}, T., {Wheatley}, P.~J., {Irwin}, P.~G.~J., {Kirk}, J., \& {Skillen},
  I. 2017, \mnras, 470, 742

\bibitem[{{Madhusudhan} {et~al.}(2020){Madhusudhan}, {Nixon}, {Welbanks},
  {Piette}, \& {Booth}}]{Madhusudhan2020}
{Madhusudhan}, N., {Nixon}, M.~C., {Welbanks}, L., {Piette}, A. A.~A., \&
  {Booth}, R.~A. 2020, \apjl, 891, L7

\bibitem[{{Malik} {et~al.}(2019){Malik}, {Kempton}, {Koll}, {Mansfield},
  {Bean}, \& {Kite}}]{Malik2019}
{Malik}, M., {Kempton}, E. M.~R., {Koll}, D. D.~B., {et~al.} 2019, arXiv
  e-prints, arXiv:1907.13135

\bibitem[{{Mallonn} {et~al.}(2019){Mallonn}, {K{\"o}hler}, {Alexoudi}, {von
  Essen}, {Granzer}, {Poppenhaeger}, \& {Strassmeier}}]{Mallonn2019}
{Mallonn}, M., {K{\"o}hler}, J., {Alexoudi}, X., {et~al.} 2019, \aap, 624, A62

\bibitem[{{Mansfield} {et~al.}(2019){Mansfield}, {Kite}, {Hu}, {Koll}, {Malik},
  {Bean}, \& {Kempton}}]{Mansfield2019}
{Mansfield}, M., {Kite}, E.~S., {Hu}, R., {et~al.} 2019, arXiv e-prints,
  arXiv:1907.13150

\bibitem[{{Maturilli} {et~al.}(2016){Maturilli}, {Helbert}, {Ferrari},
  {Davidsson}, \& {D'Amore}}]{Mat2016}
{Maturilli}, A., {Helbert}, J., {Ferrari}, S., {Davidsson}, B., \& {D'Amore},
  M. 2016, Earth, Planets, and Space, 68, 113

\bibitem[{{May} \& {Rauscher}(2016)}]{May2016}
{May}, E.~M., \& {Rauscher}, E. 2016, \apj, 826, 225

\bibitem[{{Menou} \& {Rauscher}(2009)}]{Menou2009}
{Menou}, K., \& {Rauscher}, E. 2009, \apj, 700, 887

\bibitem[{{Miguel} \& {Kaltenegger}(2014)}]{Miguel2014}
{Miguel}, Y., \& {Kaltenegger}, L. 2014, \apj, 780, 166

\bibitem[{{Miller-Ricci} \& {Fortney}(2010)}]{MR&F10}
{Miller-Ricci}, E., \& {Fortney}, J.~J. 2010, \apjl, 716, L74

\bibitem[{NEO(1999--)}]{NEO}
NEO. 1999--, NASA Earth Observatory, [Online; accessed <today>]

\bibitem[{{Ning} {et~al.}(2018){Ning}, {Wolfgang}, \& {Ghosh}}]{Ning2018}
{Ning}, B., {Wolfgang}, A., \& {Ghosh}, S. 2018, arXiv e-prints,
  arXiv:1811.02324

\bibitem[{{Perez-Becker} \& {Showman}(2013)}]{Perez2013}
{Perez-Becker}, D., \& {Showman}, A.~P. 2013, \apj, 776, 134

\bibitem[{{Pierrehumbert} \& {Hammond}(2019)}]{Pier2019}
{Pierrehumbert}, R.~T., \& {Hammond}, M. 2019, Annual Review of Fluid
  Mechanics, 51, 275

\bibitem[{{Rauscher}(2017)}]{Rauscher2017}
{Rauscher}, E. 2017, \apj, 846, 69

\bibitem[{{Rauscher} \& {Kempton}(2014)}]{Rauscher2014}
{Rauscher}, E., \& {Kempton}, E. M.~R. 2014, \apj, 790, 79

\bibitem[{{Rauscher} \& {Menou}(2012)}]{RM12}
{Rauscher}, E., \& {Menou}, K. 2012, \apj, 750, 96

\bibitem[{{Rauscher} {et~al.}(2007){Rauscher}, {Menou}, {Seager}, {Deming},
  {Cho}, \& {Hansen}}]{Rauscher2007}
{Rauscher}, E., {Menou}, K., {Seager}, S., {et~al.} 2007, \apj, 664, 1199

\bibitem[{{Rauscher} {et~al.}(2018){Rauscher}, {Suri}, \&
  {Cowan}}]{Rauscher2018}
{Rauscher}, E., {Suri}, V., \& {Cowan}, N.~B. 2018, \aj, 156, 235

\bibitem[{{Rogers}(2015)}]{Rogers2015}
{Rogers}, L.~A. 2015, \apj, 801, 41

\bibitem[{{Roman} \& {Rauscher}(2017)}]{Roman17}
{Roman}, M., \& {Rauscher}, E. 2017, \apj, 850, 17

\bibitem[{{Roman} \& {Rauscher}(2018)}]{Roman2018}
---. 2018, ArXiv e-prints, arXiv:1807.08890

\bibitem[{{Schlawin} {et~al.}(2018){Schlawin}, {Greene}, {Line}, {Fortney}, \&
  {Rieke}}]{Schlawin2018}
{Schlawin}, E., {Greene}, T.~P., {Line}, M., {Fortney}, J.~J., \& {Rieke}, M.
  2018, \aj, 156, 40

\bibitem[{{Seager} {et~al.}(2007){Seager}, {Kuchner}, {Hier-Majumder}, \&
  {Militzer}}]{Seager2007}
{Seager}, S., {Kuchner}, M., {Hier-Majumder}, C.~A., \& {Militzer}, B. 2007,
  \apj, 669, 1279

\bibitem[{{Showman} {et~al.}(2015){Showman}, {Lewis}, \&
  {Fortney}}]{Showman2015}
{Showman}, A.~P., {Lewis}, N.~K., \& {Fortney}, J.~J. 2015, \apj, 801, 95

\bibitem[{{Showman} {et~al.}(2013){Showman}, {Wordsworth}, {Merlis}, \&
  {Kaspi}}]{Showman2013}
{Showman}, A.~P., {Wordsworth}, R.~D., {Merlis}, T.~M., \& {Kaspi}, Y. 2013,
  {Atmospheric Circulation of Terrestrial Exoplanets}, ed. S.~J. {Mackwell},
  A.~A. {Simon-Miller}, J.~W. {Harder}, \& M.~A. {Bullock}, 277

\bibitem[{{Toon} {et~al.}(1989){Toon}, {McKay}, {Ackerman}, \&
  {Santhanam}}]{Toon89}
{Toon}, O.~B., {McKay}, C.~P., {Ackerman}, T.~P., \& {Santhanam}, K. 1989,
  \jgr, 94, 16287

\bibitem[{{Tsiaras} {et~al.}(2019){Tsiaras}, {Waldmann}, {Tinetti}, {Tennyson},
  \& {Yurchenko}}]{Tsiaras2019}
{Tsiaras}, A., {Waldmann}, I.~P., {Tinetti}, G., {Tennyson}, J., \&
  {Yurchenko}, S.~N. 2019, Nature Astronomy, 3, 1086

\bibitem[{{Vallis}(2006)}]{Vallis2006}
{Vallis}, G.~K. 2006, {Atmospheric and Oceanic Fluid Dynamics}, 770,
  doi:10.2277/0521849691

\bibitem[{{Washington} \& {Parkinson}(1986)}]{WPBook}
{Washington}, W.~M., \& {Parkinson}, C.~L. 1986, {An Introduction to
  Three-Dimensional Climate Modeling} (University Science Books)

\bibitem[{{Way} {et~al.}(2018){Way}, {Del Genio}, {Aleinov}, {Clune}, {Kelley},
  \& {Kiang}}]{Way2018}
{Way}, M.~J., {Del Genio}, A.~D., {Aleinov}, I., {et~al.} 2018, \apjs, 239, 24

\bibitem[{{Williams} {et~al.}(2006){Williams}, {Charbonneau}, {Cooper},
  {Showman}, \& {Fortney}}]{Williams2006}
{Williams}, P. K.~G., {Charbonneau}, D., {Cooper}, C.~S., {Showman}, A.~P., \&
  {Fortney}, J.~J. 2006, \apj, 649, 1020

\bibitem[{{Wolfgang} {et~al.}(2016){Wolfgang}, {Rogers}, \&
  {Ford}}]{Wolfgang2016}
{Wolfgang}, A., {Rogers}, L.~A., \& {Ford}, E.~B. 2016, \apj, 825, 19

\bibitem[{{Zeng} {et~al.}(2016){Zeng}, {Sasselov}, \& {Jacobsen}}]{Zeng2016}
{Zeng}, L., {Sasselov}, D.~D., \& {Jacobsen}, S.~B. 2016, \apj, 819, 127

\bibitem[{{Zeng} {et~al.}(2019){Zeng}, {Jacobsen}, {Sasselov}, {Petaev},
  {Vanderburg}, {Lopez-Morales}, {Perez-Mercader}, {Mattsson}, {Li}, {Heising},
  {Bonomo}, {Damasso}, {Berger}, {Cao}, {Levi}, \& {Wordsworth}}]{Zeng2019}
{Zeng}, L., {Jacobsen}, S.~B., {Sasselov}, D.~D., {et~al.} 2019, Proceedings of
  the National Academy of Science, 116, 9723

\bibitem[{{Zhang} {et~al.}(2017){Zhang}, {Kempton}, \& {Rauscher}}]{Zhang2017}
{Zhang}, J., {Kempton}, E. M.~R., \& {Rauscher}, E. 2017, \apj, 851, 84

\end{thebibliography}

\end{document}